# Economic Evaluation of Transformer Loss of Life Mitigation by Energy Storage and PV Generation

Milad Soleimani, Student Member, IEEE, Mladen Kezunovic, Life Fellow, IEEE,
Department of Electrical and Computer Engineering,
Texas A&M University,
College Station, Texas, U.S.A

*Abstract*—**High penetration of plug-in electric vehicles (PEVs) can potentially put the utility assets such as transformer under overload stress causing decrease in their lifetime. The decrease in PV and battery energy storage system (BESS) prices has made them viable solutions to mitigate this situation. In this paper, the economic aspect of their optimal coordination is studied to assess transformer hottest spot temperature (HST) and loss of life. Monte Carlo simulation is employed to provide synthetic data of PEVs load in a residential complex and model their stochastic behavior. For load, temperature, energy price and PV generation, real data for City of College Station, Texas, USA in 2018 is acquired and a case study is developed for one year. The results illustrate using BESS and PV is economically effective and mitigates distribution transformer loss of life.**

## NOMENCLATURE

| | |
|---|---|
| $P_{Ti,j}$ | Average power that flows through the transformer excluding the effect of battery in the last 10 days before the $j^{th}$ day in the $i^{th}$ hour |
| $P_{avg}$ | Average power that flows through the transformer in the last 10 days excluding the effect of battery |
| $P_{B_i}$ | Power of the battery, charged (+) and discharged (-) |
| $\eta$ | Efficiency of the battery and inverter |
| $E_C$ | Capacity of the battery in KWh |
| $t_{C_1}, t_{C_2}$ | Starting and ending time of charging |
| $t_{D_1}, t_{D_2}$ | Starting and ending time of discharging |
| $SOC_t$ | Battery state of charge at time t |
| $SOC^{PEV}$ | PEV battery state of charge |
| $E_D^{PEV}$ | Electricity consumption in KWh/mile |
| $T_{CH}^{PEV}$ | Required time to charge PEV |
| $\Theta_H$ | Hottest spot temperature |
| $\Theta_a$ | Ambient temperature |
| $\Delta\theta_{to}$ | Top oil temperature rise over the ambient |
| $\Delta\theta_h$ | Winding hottest spot temperature rise over the top oil |
| $K_u$ | Ratio of ultimate load to the expected load |
| $R$ | Ratio between loss at rated load and no load |
| $F_{AA}$ | Aging acceleration factor |
| $F_{EQA}$ | Equivalent aging factor |
| $TP$ | Time period under analysis in hour |
| $L$ | Normal insulation life |
| $\tau_\omega, \tau_{to}$ | Winding and oil time constant |
| $ini$ | Refers to the initial value |
| $ult$ | Refers to the ultimate value |
| $Inv_0$ | Initial investment |
| $PR_t$ | Profit at time interval t |
| $r$ | Discount rate |
| $PB$ | Payback period |

## I. INTRODUCTION

Plug-in Electric vehicles (PEVs) have been considered as a promising solution to reduce fossil fuel consumption and greenhouse gas emission. The PEV global market is rising. Thus, in the near future, power system will face a new challenge from high penetration of PEVs. It is challenging since it is expected that a large number of PEVs will be connected to get charged in residential buildings with considerable overlap in charging time. This will expose assets such as transformers to overloading stress.

Increasing presence of PEVs will increase the electricity consumption and this increase can cause several issues such as power quality deterioration [1], under-voltage condition [2] and extra demand on distribution transformers [3]. Usually distribution transformers do not have on-line monitoring and when they operate in an overload condition continuously, they will face accelerated aging and risk of their failure will increase [4]. In addition to an unexpected outage, their more frequent failure will cause increased costs of repair or replacement.

Battery energy storage system (BESS) cost has reduced and paved the path to be considered as a promising solution to mitigate the negative impacts of PEVs' high penetration. BESS

This material is based upon work supported by the Department of Energy, Office of International Affairs and Office of Electricity under Award Number(s) DE-IA0000025.

can be employed for different applications such as energy arbitrage [5], frequency and voltage support [6], peak shaving [7] and congestion management [8]. One of the additional benefits of using ES, considered in this paper, can be reducing the accelerated aging of the transformers.

The adverse effect of PEVs on the transformers is studied in [9] - [11]. Reference [3] illustrates how different level of PEV penetration will negatively affect the transformers in a residential complex. In [12], a probabilistic approach is proposed to investigate the impact of EV on transformers loss of life when PV generation is present.

Reference [10] is the only one among the abovementioned papers that takes stationary battery energy storage into account. However, the main focus is proposing a smart charging approach in a presence of a predetermined BESS. Moreover, the economic analysis in [10] employs a dynamic pricing method for the customers and schedules BESS charging/discharging by optimizing the price for the customers. The contribution of our paper is studying the economic benefits of employing BESS and PV generation considering the mitigation of transformer loss of life that has not been studied before and yet it is shown that it is essential to consider it.

For this purpose, one year data of PEVs demand is created using a probabilistic approach. Load [13], irradiation [14], temperature [15] and electricity price data in the city of College Station, Texas, USA in 2018 are used to provide a realistic evaluation. The payback period of different scenarios is calculated and it is shown that employing BESS and PV generation yields tangible financial benefits.

The remainder of this paper is organized as follows. Section II shows the battery scheduling methodology. In section III, the utilized data is introduced. In Section IV, the employed method for generating PEVs demand is illustrated. Section V presents the transformer aging model. In Section VI, the utilized economic model is proposed. In Section VII, the results are demonstrated and discussed and finally, in Section VIII, the main conclusions are outlined, followed by list of references.

II. BATTERY CHARGING/DISCHARGING OPTIMIZATION

The residential building that is used in this paper is shown in Figure 1. The nominal power of the transformer is 63KVA and PV system size is 10KW. The building load is connected to the same bus as PV panel, battery storage, and charging station. It is assumed that PEV can only be charged, i.e. it operates in the grid to vehicle mode. The fixed BESS operates both in charging and discharging modes. The charging station is located in the parking lot of the residential building.

In this paper, the assumed role of the battery is peak load shaving. It is assumed that the battery scheduling is day-ahead and the minimum state of charge is 20%. To prevent battery accelerated ageing, one cycle a day maximum is considered for battery charging/discharging. The following is total power flowing through the transformer in the absence of BESS during ten consecutive days before the current day.

$$P_{T_{i,j}} = \sum_{k=1}^{10} \frac{(P_{Load_{i,j-k}} + P_{PEV_{i,j-k}} - P_{PV_{i,j-k}})}{10} \quad (1)$$

For simplification, index j in further equations is removed. The optimization problem for the current day (j[th] day) is formulated as follows:

$$Min\ C = \sum_{i=1}^{24} |P_{T_i} + P_{B_i} - P_{avg}| \quad (2)$$

The battery should not be charged more than 100% and should not be discharged to less than 20% of selected state of charge. Hence, constraints (3) and (4) can be defined as follows:

$$\sum_{i=t_{C1}}^{t_{C2}} P_{B_i} \leq \eta E_C (1 - SOC_{t_{C1}}) \quad (3)$$

$$\sum_{i=t_{D1}}^{t_{D2}} -P_{B_i} \leq \eta E_C (SOC_{t_{C3}} - 0.2) \quad (4)$$

Furthermore, the charging or discharging power should not be more than the nominal power of the battery charger that leads to the constraint (5).

$$|P_{B_i}| \leq P_n \quad (5)$$

The starting time of charging and discharging must be positive and should not exceed the ending time. The problem should obey the following constraints (6) – (8).

$$t_{C1} \leq t_{C2} \quad (6)$$

$$t_{D1} \leq t_{D2} \quad (7)$$

$$t_{C1}, t_{C2}, t_{D1}, t_{D2} \geq 0 \quad (8)$$

Constraint (9) and (10) are defined for when charging is scheduled before discharging and when discharging is scheduled before charging, respectively.

$$If\ t_{C1} < t_{D1} \Rightarrow SOC_{t_{D1}} = SOC_{t_{C1}} + \frac{\sum_{i=t_{C1}}^{t_{C2}} P_{B_i}}{\eta E_C} \quad (9)$$

$$If\ t_{C1} > t_{D1} \Rightarrow SOC_{t_{C1}} = SOC_{t_{D1}} + \frac{\sum_{i=t_{D1}}^{t_{D2}} P_{B_i}}{\eta E_C} \quad (10)$$

Since the planning is day-ahead, heuristic optimization methods can be utilized to solve this optimization problem. In this paper, genetic algorithm is deployed.

III. LOAD, SOLAR GENERATION AND TEMPERATURE DATA

A. Load Data

The load data is acquired from National Renewable Energy Laboratory (NREL), OpenEI [13]. The dataset contains the hourly load profile for different types of commercial and residential buildings. The data for the residential buildings is synthesized based on Building America House Simulation Protocols [16] and Residential Energy Consumption Survey (RECS) [17]. In the under study, there are 10 total customers from which energy consumption of 3 is low, 6 is medium and 1 is high.

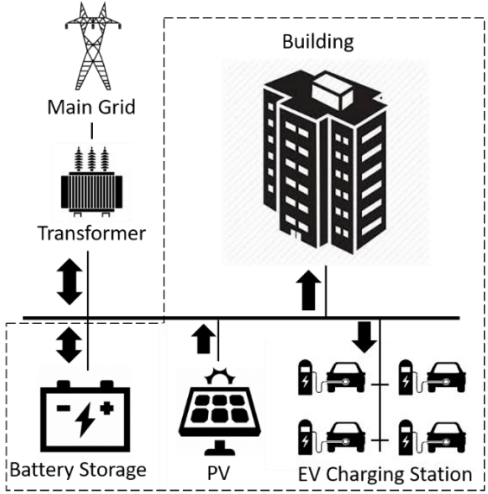

Figure 1. Schematic of the system under study.

### B. Solar Generation Data

PVWatts Calculator [18] is employed to calculate the hourly generation of the PV considering panel, inverter and solar irradiation. This system employs the newest data from National Solar Radiation Database and by receiving the information of the PV panel, can provide hourly data of PV generation for a specific location.

### C. Temperature Data

The hourly temperature data is obtained from Iowa Environmental Mesonet [19]. The temperature data are collected using Automated Surface Observing Systems. The temperature of each hour is assumed to be fixed.

## IV. MODELLING UNCERTAINTIES OF PEV CHARGING DEMAND

To model the uncertainty of the demand imposed by PEVs, the approach introduced in [9], [10] is used. In order to estimate PEV load profile several variables such as driving distance, arrival time and PEVs' characteristics are considered. The flowchart of the estimation process is shown in Figure 2.

In a residential building, it is more probable that PEV owners leave the building in the morning and return in the evening. Thus, the following distribution function is used. The time of arrival is modeled with a normal distribution with the mean value of $\mu_d$ = 17:00 and standard deviation of $\sigma_d$ = 2.28.

$$AT = Gauss(\mu_{AT}, \sigma_{AT}^2) \quad (11)$$

The daily driving distance is modeled using log-normal distribution with the mean value of $\mu_d$ = 3.37 and standard deviation of $\sigma_d$ = 0.5. The distribution is defined as follows:

$$d = Ln(\mu_d, \sigma_d^2) \quad (12)$$

The SOC when PEV is driven home can be calculated using equation (13):

$$SOC_{ini}^{PEV} = 1 - \frac{E_D^{PEV} \cdot d}{E_C} \quad (13)$$

It is assumed that the required SOC is 0.95. Thus, considering the PEVs are charged with constant power, the required time to charge the vehicle can be calculated as follows:

$$T_{CH}^{PEV} = \frac{(SOC_{req}^{PEV} - SOC_{ini}^{PEV}) \cdot E_C^{PEV}}{\eta^{PEV} \cdot P} \quad (14)$$

Nissan Leaf is the studied electric vehicle in this paper. The battery capacity in this car is 24KWh and it consumes 340Wh/mile. It is assumed the residents own 12 PEVs cars and there are 10 charging slots in the building parking area.

## V. TRANSFORMER AGEING

Loading and ambient temperature are the main factors that affect hot spot temeperature (HST) and transformer loss of life. When transformer is working under overload stress, the temperature adjacent to the winding of transformer increases and deteriorates winding insulation. It causes accelerated ageing for the winding insulation and decreases the life of the asset. IEEE Standard C57.91 [20] thoroughly explains quantification of this effect on loss of life. Firstly, the hottest spot temperature can be calculated:

$$\theta_H = \theta_a + \Delta\theta_{to} + \Delta\theta_h \quad (15)$$

For which the changes in temperature can be calculated as follows:

$$\Delta\theta_h = (\Delta\theta_{h,ult} - \Delta\theta_{h,ini}) \cdot (1 - e^{-\frac{t}{\tau_w}}) + \Delta\theta_{h,ini} \quad (16)$$

$$\Delta\theta_{to} = (\Delta\theta_{to,ult} - \Delta\theta_{to,ini}) \cdot (1 - e^{-\frac{t}{\tau_{to}}}) + \Delta\theta_{to,ini} \quad (17)$$

$$\Delta\theta_{h,u} = \Delta\theta_{h,rated\_load} \cdot K_u^{2m} \quad (18)$$

$$\Delta\theta_{to,ult} = \Delta\theta_{to,rated\_Load} \cdot \left[\frac{K_u R + 1}{R + 1}\right]^n \quad (19)$$

For which m and n are empirically derived exponents and their values depend on the cooling of transformer, the aging acceleration factor and the equivalent aging factor are $F_{AA}$ and $F_{EQA}$. They can be calculated as follows:

$$F_{AA}(t) = e^{(\frac{15000}{110+273} - \frac{15000}{\theta_H + 273})} \quad (20)$$

$$F_{EQA} = \frac{\sum_{n=1}^{N} F_{AA,n} \Delta t_n}{\sum_{n=1}^{N} \Delta t_n} \quad (21)$$

And loss of life can be calculated using the following equation:

$$Loss\ of\ Life = \frac{F_{EQA} \cdot TP}{L} \quad (22)$$

## VI. ECONOMIC MODEL

PV and BESS are relatively expensive equipment and it is essential to evaluate the viability of deploying them. Due to energy loss in the charging/discharging procedure, it is significant to evaluate the economic benefits. Several energy pricing and selling methods to residential customers exist and this may vary in different areas or cities. In the College Station area, which is the subject of study in this paper, the customers buy electricity from College Station Utility with fixed prices. In addition to a $7 monthly service fee, the electricity price is $0.1369/KWh from May to October and $0.1323/KWh from November to April including tax and transmission fees [21]. The College Station Utility buys the electricity based on the prices announced by Electric Reliability Council of Texas (ERCOT) every 5 minutes.

The utilities are usually the owner of distribution transformers and they are the beneficiary of loss of life mitigation effect of PV and BESS utilization. Considering the electricity price, it is clear that the beneficiary of employing BESS is the utility and not customers, but utilities do not necessarily benefit from PV generation. In this context, customers should receive incentives from utilities to make using BESS economically attractive for them. Incentive determination is out of scope of this paper and the profit for both customer and utility is calculated to obtain the payback period of the investment regardless which parties share the investment.

Net present value (NPV), shown in (23), is the difference between the present value of cash inflows and the present value of cash outflows over a period of time i.e. it is the difference between the current value of all incomes and investments. NPV is a usual index in investment planning to analyze the profitability of a project or investment. The payback (PB) period is the required time to recover the initial investment and can be obtained using (24).

$$NPV = -Inv_0 + \sum_{t=0}^{T} \frac{PR_t}{1+r} \quad (23)$$

$$if \ \exists T_{PB} : \forall T > T_{PB}, NPV \geq 0 \Rightarrow PB = T_{PB} \quad (24)$$

Table I shows the summary of the parameters values used in the economic analysis. Transformer remaining life is assumed to be 112000 hours out of 180000 hours for a new one. Information regarding costs and rates are acquired using different available sources [22] – [23]. Energy price increase rate is taken as the average of this value in USA in the last 20 years. Used discount rate is announced by the U.S. Federal Reserve Board of Governors effective December 2018.

TABLE I. COSTS AND FINANCIAL PARAMETERS

| Equipment Costs | | Financial Parameters | |
|---|---|---|---|
| PV System | $2.13/W | Energy price increase rate | 2.6% /year |
| BESS | $0.4/W | | |
| Transformer NPV | $5000 | Discount Rate | 3.0% /year |

To include transformer loss of life in the economic calculations equivalent annual cost (EAC) is used. EAC is the cost per year of owning and operating an asset over its entire lifespan and is calculated using the equation (25). In this equation, t is the expected life of the asset and NPV is the net present value of that.

$$EAC = \frac{NPV_{transformer} \cdot r}{1 - \frac{1}{(1+r)^t}} \quad (25)$$

The actual extra cost caused by transformer loss of life is the difference between EAC before investment on BESS and/or PV and after that. Therefore, PB can be calculated using equations (24) and (26).

$$NPV = -Inv_0 + \sum_{t=0}^{T}(\frac{PR_t}{(1+r)^t} + EAC_2 - EAC_1) \quad (26)$$

## VII. SIMULATION RESULTS

The method is implemented in MATLAB, the transformer loss-of-life evaluation and economic calculations is performed for the following scenarios.

a) No PV generation, no PEV, and no energy storage.
b) No PV generation, high penetration of PEVs and no energy storage.
c) PV generation, high penetration of PEVs and no energy storage.
d) No PV generation, high penetration of PEVs and several levels of energy storage.
e) PV generation, high penetration of PEVs and several levels of energy storage.

The results of 7 cases are shown in Table II. In this table, it can be clearly seen how high penetration of PEVs can negatively impact the life of the transformer and the effect can be mitigated using PV generation and BESS.

TABLE II. TRANSFORMER LOSS OF LIFE FOR DIFFERENT SCENARIOS

| Scenario | Transformer Loss of Life (%) |
|---|---|
| a | 0.0124 |
| b | 33.45 |
| c | 12.23 |
| d.a- d and the capacity of the battery is 20KWh. | 6.86 |
| d.b- d and the capacity of the battery is 40KWh. | 1.6 |
| e.a- e and the capacity of the battery is 20KWh. | 2.3125 |
| e.b- e and the capacity of the battery is 40KWh. | 0.45 |

Using the data, simulation results and the introduced economic model, payback periods of the abovementioned scenarios, for when loss of life is considered and when it is not, are calculated and shown in Table III.

In Table III, it is shown how considering the economic impact of transformer loss of life can dramatically change the economic calculations. The conspicuous difference is in scenario d for which only BESS is employed. If the effect of BESS in transformer loss of life is not taken into account in the economic evaluations, utilizing BESS may look an unfeasible solution, but when its impact on transformer life is considered, it shows a promising solution to overcome the changes in the load curve for future grid with high penetration of PEVs. Total annual saving is shown in Table III.

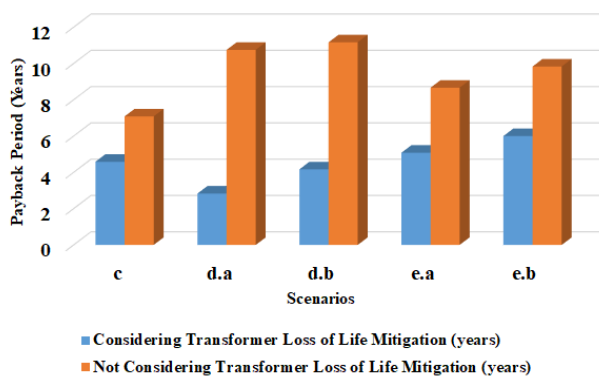

Figure 2. Payback period for different scenarios

TABLE III. TOTAL ANNUAL SAVINGS FOR DIFFERENT SCENARIOS

| Scenario | c | d.a | d.b | e.a | e.b |
|---|---|---|---|---|---|
| Saving ($) | 4499.3 | 2813.24 | 3835.84 | 5597.22 | 6029.58 |

## VIII. CONCLUSION

We have studied how the impact of BESS and PV generation on transformer loss of life can vary the economic evaluation of the total investment by simulating a residential building demand for different scenarios of PV generation and PEV charging penetration. Specifically, we have shown that ignoring positive effect of BESS on transformer's life can be misleading. The demand for PEV charging, generated using a probabilistic method based on data for the city of College Station, TX. The results show that:

- Inclusion of 12 PEVs and 10 parking spaces with charging stations in a building with 10 residents will greatly accelerate the aging of the transformer and reduce its life.
- An EAC economic model of the costs of transformer loss of life shows that if only the profit from BESS energy storage is considered, payback period will be rather long
- Increasing the capacity of battery or using both PV generation and BESS may make the payback period longer since the value of mitigation does not offset the cost of investment.
- Utilizing the proposed approach for optimal use of PV generation and BESS will mitigate the transformer loss of life caused by PEVs, and at the same time the payback period is short enough making the arrangement not only viable but also profitable.